\pdfoutput=1
\documentclass{PoS}

\usepackage{xspace}
\usepackage{subfig}
\usepackage{multirow}

\newcommand*{\POWHEGBOX}{\textsc{Powheg-Box}\xspace}
\newcommand*{\SHERPA}{\textsc{Sherpa}}
\newcommand*{\PYTHIA}{\textsc{Pythia}}
\newcommand*{\MGMCatNLO}{\textsc{MadGraph5}\_aMC@NLO\xspace}
\newcommand*{\antibar}[1]{\ensuremath{#1\bar{#1}}\xspace}
\newcommand*{\ttbar}{\antibar{t}}
\newcommand*{\TeV}{\ifmmode {\mathrm{\ Te\kern -0.1em V}}\else
                   \textrm{Te\kern -0.1em V}\fi}%
\newcommand*{\GeV}{\ifmmode {\mathrm{\ Ge\kern -0.1em V}}\else
                   \textrm{Ge\kern -0.1em V}\fi}%
\newcommand*{\pt}{\ensuremath{p_{\rm T}}\xspace}
\newcommand*{\pT}{\ensuremath{p_{\rm T}}\xspace}
\newcommand*{\HT}{\ensuremath{H_{\mathrm{T}}}\xspace}
\newcommand*{\met}{\ensuremath{E_{\mathrm{T}}^{\mathrm{miss}}}\xspace}
\newcommand{\coll}[1]{#1 Collaboration}
\newcommand{\apub}[3]{ATLAS-PHYS-PUB-#1-#2, 
\href{https://cds.cern.ch/record/#3}{\color[rgb]{0.,0.7,0.}
{https://cds.cern.ch/record/#3}}}
\newcommand{\arxiv}[2]{arXiv:\href{http://www.arxiv.org/abs/#1}{\color[rgb]{0.,0.7,0.}{#1 [hep-#2]}}}
\newcommand{\pub}[2]{\href{http://dx.doi.org/#2}{\color[rgb]{0.,0.7,0.}{#1}}}
\newcommand{\accept}[1]{accepted for publication in #1}

\newcommand{\PRD}{Phys.\ Rev.\ D}
\newcommand{\EPJC}{Eur.\ Phys.\ J.\ C}
\newcommand{\PLB}{Phys.\ Lett.\ B}
\newcommand*{\WWWs}{\ensuremath{W^{\pm} W^{\pm} W^{\mp}}\xspace}
\newcommand*{\WWW}{\ensuremath{WWW}\xspace}
\newcommand*{\WWZs}{\ensuremath{W^{\pm} W^{\mp} Z}\xspace}
\newcommand*{\WWZ}{\ensuremath{WWZ}\xspace}
\newcommand*{\WZZs}{\ensuremath{W^{\pm} Z Z}\xspace}
\newcommand*{\WZZ}{\ensuremath{WZZ}\xspace}
\newcommand*{\WVZ}{\ensuremath{WVZ}\xspace}
\newcommand*{\ttZ}{\ensuremath{t\bar{t}Z}\xspace}
\newcommand{\sswww}{\ensuremath{WWW \to \ell \nu \ell \nu qq}\xspace}
\newcommand{\lllwww}{\ensuremath{WWW \to \ell \nu \ell\nu \ell \nu}\xspace}
\newcommand{\lllwvz}{\ensuremath{WVZ \to \ell \nu qq \ell \ell}\xspace}
\newcommand{\llllwwz}{\ensuremath{WWZ \to \ell \nu \ell \nu \ell \ell}\xspace}
\newcommand{\llllwzz}{\ensuremath{WZZ \to qq \ell \ell \ell \ell}\xspace}
\newcommand{\lljj}{\ssll}
\newcommand{\lllv}{\threelep}
\newcommand{\elel}{\ensuremath{ee}\xspace}
\newcommand{\elmu}{\ensuremath{e\mu}\xspace}
\newcommand{\muel}{\ensuremath{\mu e}\xspace}
\newcommand{\mum}{\ensuremath{\mu\mu}\xspace}
\newcommand{\ssll}{\ensuremath{\ell \nu \ell \nu qq}\xspace}
\newcommand{\threelep}{\ensuremath{\ell \nu \ell \nu \ell \nu}\xspace}
\newcommand*{\TLjone}{3$\ell$-1j\xspace}
\newcommand*{\TLjtwo}{3$\ell$-2j\xspace}
\newcommand*{\TLjthree}{3$\ell$-3j\xspace}
\newcommand*{\FLDF}{4$\ell$-DF\xspace}
\newcommand*{\FLSFZ}{4$\ell$-SF-Z\xspace}
\newcommand*{\FLSFnoZ}{4$\ell$-SF-noZ\xspace}
\newcommand*{\invfb}{\ensuremath{\textrm{fb}^{-1}}\xspace}
\newcommand*{\lumi}{\ensuremath{79.8\,\invfb\xspace}}
\newcommand*{\obsvvverr}{\ensuremath{1.40^{+0.39}_{-0.37}}\xspace}
\newcommand{\plt}{non-prompt lepton BDT}
\newcommand{\cft}{charge misidentification suppression BDT}
\newlength{\figwidth}
\title{Evidence for the production of three massive vector bosons with the ATLAS detector}

\ShortTitle{Production of three massive vector bosons}

\author{\speaker{Markus Cristinziani}\thanks{Supported by the European Research
Council grant ERC--CoG--617185 and by the German Federal Ministry of Education and
Research (FSP-103)}\\
{\rm On behalf of the ATLAS Collaboration}\\
Physikalisches Institut, Universit\"at Bonn, Nussallee 12, 53115 Bonn, Germany.\\
E-mail: \email{cristinz@uni-bonn.de}}

\abstract{%
The search for the production of three massive vector bosons in proton--proton
collisions, performed using data at $\sqrt{s} = 13\,\TeV$ recorded with the
ATLAS detector at the Large Hadron Collider in the years 2015--2017,
corresponding to an integrated luminosity of $79.8$ fb$^{-1}$, is presented. 
Events with two
same-sign leptons $\ell$ (electrons or muons) and at least two reconstructed
jets are selected to search for $WWW \to \ell \nu \ell \nu qq$. Events with
three leptons without any same-flavour opposite-sign lepton pairs are used to
search for $WWW \to \ell \nu \ell\nu \ell \nu$, while events with three leptons
and at least one same-flavour opposite-sign lepton pair and one or more
reconstructed jets are used to search for $WWZ \to \ell \nu qq \ell \ell$.
Finally, events with four leptons are analysed to search for $WWZ \to \ell \nu
\ell \nu \ell \ell$ and $WZZ \to qq \ell \ell \ell \ell$. Evidence for the
joint production of three massive vector bosons is observed with a significance
of 4.1 standard deviations, where the expectation is 3.1 standard deviations.}

\FullConference{7th Annual Conference on Large Hadron Collider Physics --- LHCP2019\\
		20--25 May, 2019\\
		Puebla, Mexico}

\begin{document}

\section{Introduction}

The joint production of three vector bosons is a rare process in the Standard
Model (SM).  Studies of triboson production can test the non-Abelian gauge
structure of the SM theory and any deviations from the SM prediction would
provide hints of new physics at higher energy scales.  Triboson production has
been studied at the Large Hadron Collider (LHC) using proton--proton ($pp$)
collision data taken at $\sqrt{s}= 8\,\TeV$ for processes involving at least
one photon, and for the \WWW process~\cite{STDM-2015-07}. 

\begin{figure}[htbp]
\centering
\includegraphics[width=0.23\figwidth]{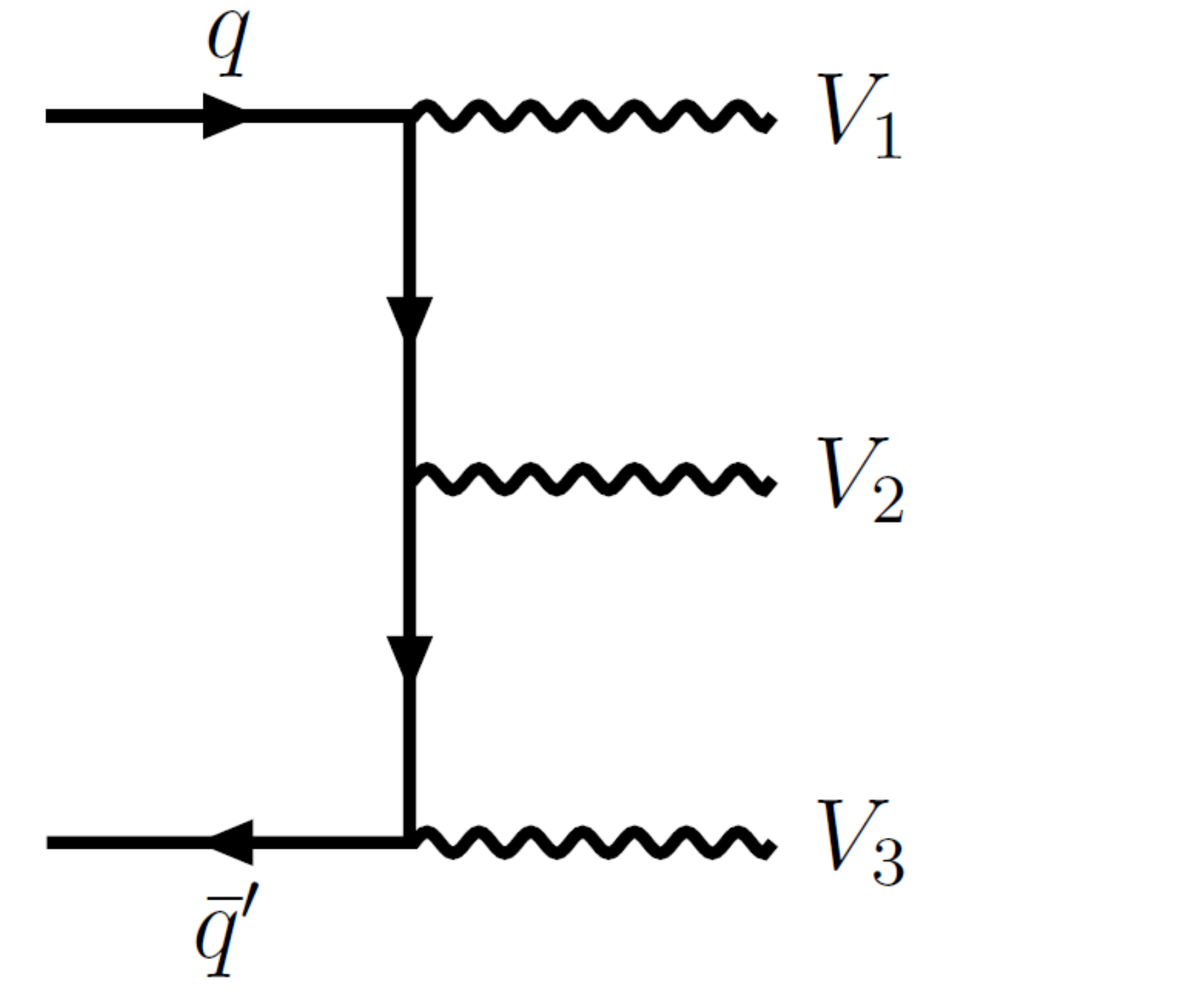} \hspace{0.05em}
\includegraphics[width=0.28\figwidth]{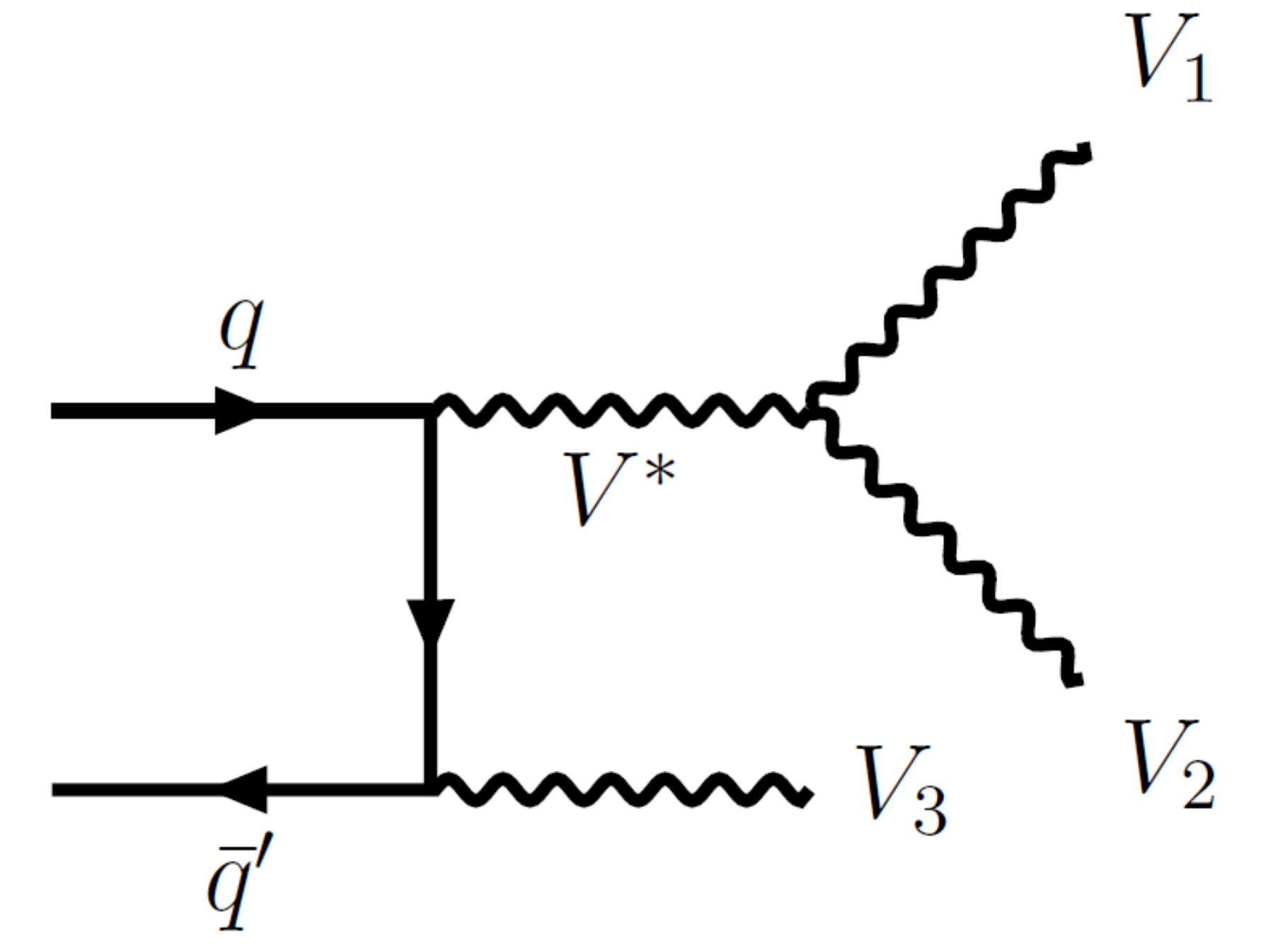} \hspace{0.5em}
\includegraphics[width=0.26\figwidth]{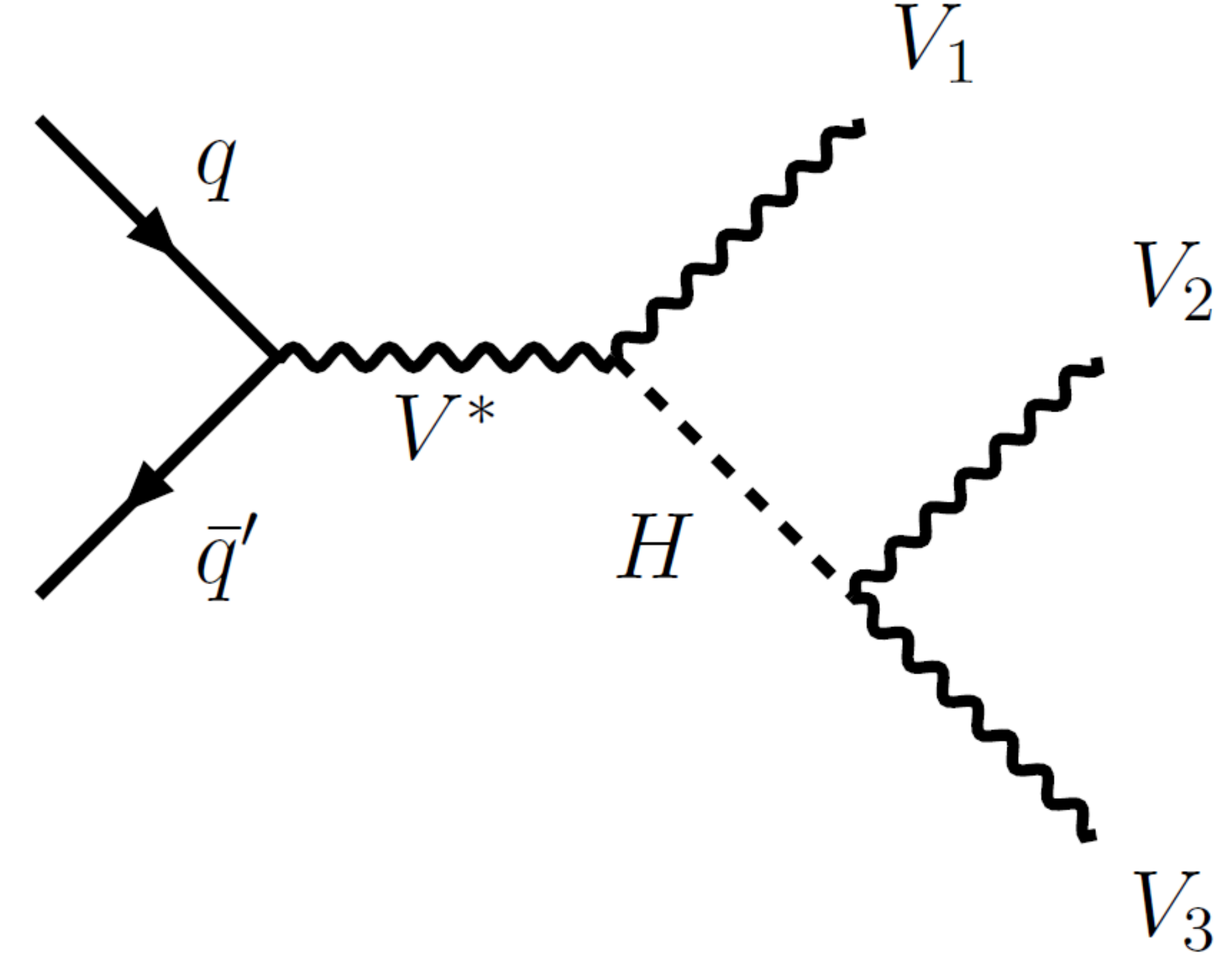} \hspace{0.5em}
\includegraphics[width=0.28\figwidth]{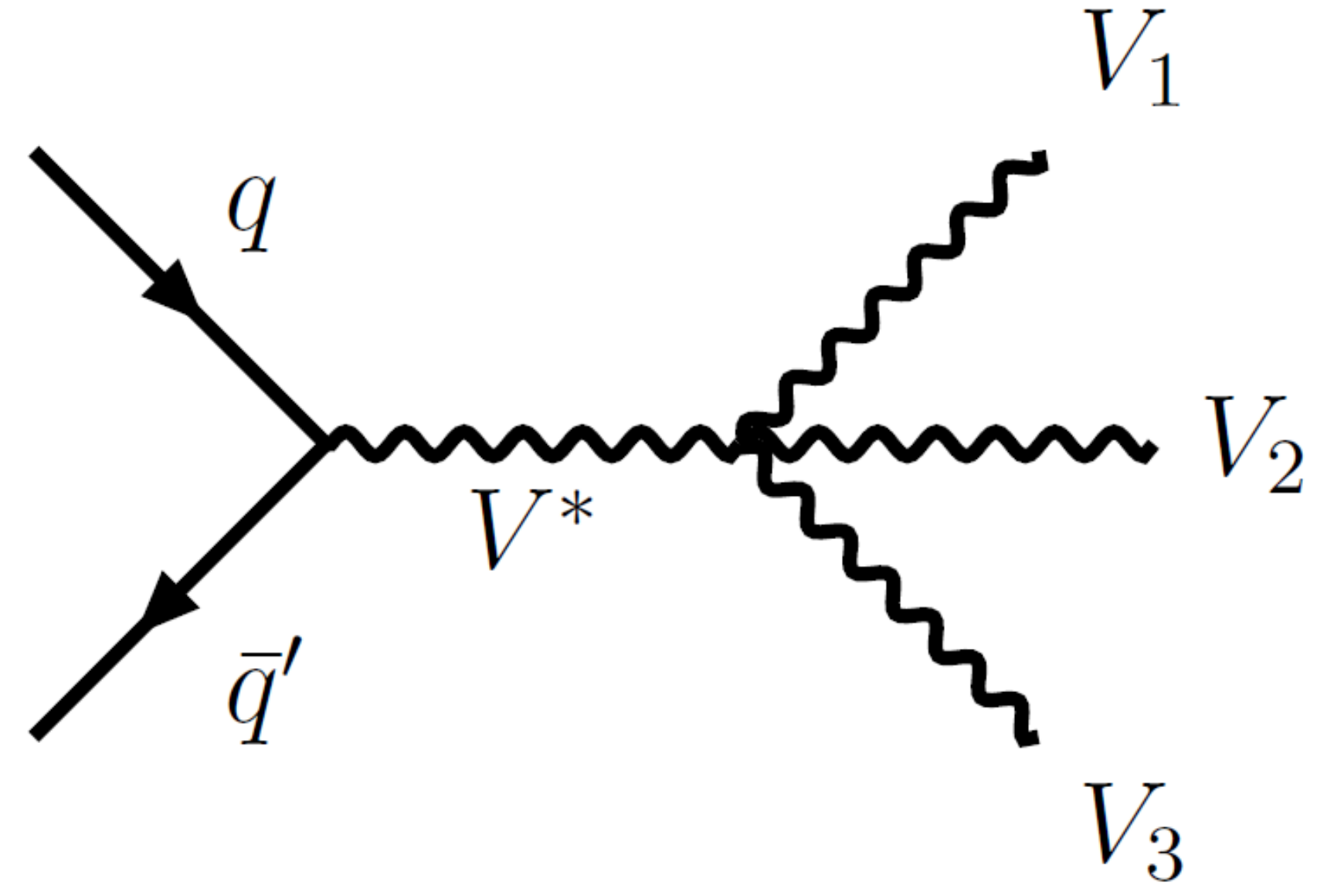}
\caption{Representative Feynman diagrams at LO for the production of three
massive vector bosons, including diagrams sensitive to triple and quartic gauge
couplings.}
\label{fig:feyn}
\end{figure}

Here, the first evidence for the joint production of three massive vector
bosons in $pp$ collisions using the dataset collected with the ATLAS
detector~\cite{atlas} between 2015 and 2017 at $\sqrt{s} =
13\,\TeV$~\cite{WVV}, for a total integrated luminosity of \lumi, is presented.
At leading order (LO) in QCD, the production of three massive vector bosons
($VVV$, with $V=W, Z$) can proceed via the radiation of each vector boson from
a fermion, from an associated boson production with an intermediate boson ($W$,
$Z/\gamma^*$ or $H$) decaying into two vector bosons, or from a quartic gauge
coupling vertex.  Representative Feynman diagrams are shown in
Figure~\ref{fig:feyn}.

\section{Analysis strategy}

Two dedicated searches are performed, one for the \WWWs (denoted as \WWW)
process and one for the \WWZs (denoted as \WWZ) and \WZZs (denoted as \WZZ)
processes.  To search for the \WWW process, events with two same-sign leptons
with at least two jets resulting from \sswww ($\ell=e, \mu$, including $\tau
\to \ell \nu \nu$) or three leptons resulting from \lllwww are considered and
are hereafter referred to as the \lljj and \lllv channels, respectively.  To
search for the \WWZ and \WZZ (denoted as \WVZ) processes, events with three or
four leptons resulting from \lllwvz, \llllwwz, and \llllwzz are used. Selection
criteria are chosen in order to ensure there is no overlap between different
channels.  A discriminant that separates the \WWW or \WVZ signal from the
background is defined in each channel.  The discriminants are combined using a
binned maximum-likelihood fit, which allows the signal yield and the background
normalisations to be extracted.  The combined observable is the signal strength
parameter $\mu$ defined as the ratio of the measured $WVV$ cross section to its
SM expectation, where one common ratio is assumed for $WWW$ and $WVZ$.

Signal and background processes were simulated with several Monte Carlo (MC)
event generators. Triboson signal events~\cite{vvvpub1} were generated using
\SHERPA, where all three bosons are on-mass-shell, using a factorised
approach~\cite{vvvpub2}.  Events with an off-mass-shell boson through $WH
\rightarrow WVV^*$ and $ZH \rightarrow ZVV^*$ were generated using \POWHEGBOX
for the $WWW$ analysis, while for the \WVZ analysis \PYTHIA\ was used.  Both
on-mass-shell and off-mass-shell processes were generated at next-to-leading
order (NLO) QCD accuracy and are included in the signal definition. 

Diboson ($WW$, $WZ$, $ZZ$), $W/Z+\gamma$ and single boson ($W/Z$+jets)
production, as well as electroweak production of $W^\pm W^\pm+2$ jets, $WZ+2$
jets, and $ZZ+2$ jets, were modelled using \SHERPA.  In order to improve the
agreement between the simulated and observed jet multiplicity distributions for
the $WZ \rightarrow \ell \nu \ell \ell$ and $ZZ \rightarrow \ell \ell \ell
\ell$ events, a jet-multiplicity based reweighting was applied to the simulated
$WZ$ and $ZZ$ samples.  Top-quark pair events (\ttbar) and other background
processes containing top quarks were generated using \POWHEGBOX or \MGMCatNLO.

Leptons are required to pass certain identification quality requirements and to
be isolated from other particles in both the calorimeters and the inner
detector.  The lepton isolation cone size is at most $\Delta R = 0.2$, except
for the muon isolation in the inner detector, where it is at most $\Delta R =
0.3$.  The requirements are more restrictive in the \WWW analysis because a
larger contamination is expected from jets misidentified as leptons or leptons
from hadron decays (including $b$- and $c$-hadron decays), referred to as
``non-prompt'' leptons in the following.

A dedicated boosted decision tree (BDT), termed ``\plt''~\cite{plt}, is used to
reject leptons likely to originate from heavy-flavour decays.  In addition,
electrons have to pass the ``\cft''~\cite{cft} to reject electrons likely to
have the electric charge wrongly measured. 

\section{Analysis targeting \WWW}

The experimental signature of the \lljj process is the presence of two
same-sign leptons, \met, and two jets. The signature of the \lllv process is
the presence of three leptons and \met.  To reduce the background contributions
from processes that have more than two (three) leptons in the \ssll (\threelep)
channel a ``veto lepton'' definition is introduced.  Compared with the nominal
lepton selection criteria, the veto lepton \pt threshold is lowered, and the
isolation, \plt, \cft, and impact parameter requirements are removed. 

To select \ssll candidates, events are required to have exactly two nominal
leptons and the same electric charge, at least two jets, and no identified
$b$-jets. Four regions are considered, based on the lepton flavour, namely
\elel, \elmu, \muel, and \mum.  The invariant mass of the dilepton system is
required to be in the range $40 < m_{\ell\ell} < 400\,\GeV$.  The upper mass
limit reduces the contribution from the $WZ$+jets process.  A dijet system,
formed by the two jets with the largest \pT, is required to have $m_{jj} <
300\,\GeV$ and $|\Delta \eta_{jj}| < 1.5$.  The cuts applied on the dijet
system mainly reduce the contributions from the same-sign $WW$ vector boson
scattering process.  Additionally, in the \elel final state, \met is required
to be above $55\,\GeV$ and $m_{\ell\ell}$ must satisfy $m_{\ell\ell} <
80\,\GeV$ or $m_{\ell\ell} > 100\,\GeV$, to reduce contamination from $Z \to
ee$ where the charge of one electron is misidentified.  To select \threelep
candidates, events are required to have exactly three nominal leptons and no
identified $b$-jets.  To reduce the contribution from the $WZ \to \ell \nu \ell
\ell$ process, events are required to have no same-flavour opposite-sign (SFOS)
lepton pairs.

A major background originates from the $WZ$+jets $\to \ell \nu \ell \ell$+jets
process, contributing to the \ssll channel when one lepton is not reconstructed
or identified, or to the \threelep channel when a $Z$ boson decays into a pair
of $\tau$ leptons both of which decay to an electron or muon.  Simulation is
used to estimate this background.  Data and simulation agree in a dedicated
validation region.  Contributions from SM processes that produce at least one
non-prompt lepton are estimated using a data-driven method as described in
Ref.~\cite{fakefactor}, where data control regions are scaled by a ``fake
factor'', which is derived from two \ttbar-enriched regions selected with two
or three leptons (no SFOS lepton pairs) and exactly one $b$-jet.  Events
resulting from the $V\gamma jj$ production can pass the \elel, \elmu and \muel
signal selection criteria if the photon is misreconstructed as an electron.
This contribution (referred to as ``$\gamma$ conv.'') is evaluated using a
data-driven method similar to the non-prompt lepton background evaluation by
introducing ``photon-like'' electrons.  The charge misidentification background
originates from processes that produce oppositely-charged prompt leptons, where
one lepton's charge is misidentified and results in final states with two
same-sign leptons.  The background is estimated using a data-driven technique.

\section{Analysis targeting \WWZ and \WZZ}
\label{s:wvz}

The experimental signature of the \lllwvz, \llllwwz, and \llllwzz processes is
the presence of three or four charged leptons.  In order to increase the signal
acceptance, ``loose'' leptons are defined in addition to nominal leptons, the
latter being a subset of the former.  Loose leptons have both the isolation and
\plt\ requirements removed.  In addition, for loose electrons the \cft\
requirement is removed.  Six regions are defined with either three or four
loose leptons, sensitive to triboson final states containing $Z$ bosons.  Among
all possible SFOS lepton pairs, the one with $m_{\ell\ell}$ closest to the $Z$
boson mass is defined as the best $Z$ candidate.  In all regions, the presence
of such a best $Z$ candidate with $|m_{\ell\ell} - 91.2\,\GeV| < 10\,\GeV$, is
required.  Furthermore, any SFOS lepton pair combination is required to have a
minimum invariant mass of $m_{\ell\ell} > 12\,\GeV$.  Events with $b$-tagged
jets are vetoed.

For the three-lepton channel, the lepton which is not part of the best $Z$
candidate is required to be a nominal lepton. The scalar sum of the transverse
momenta of all leptons and jets (\HT) is required to be larger than
$200\,\,\GeV$. This significantly reduces the contribution of the
$Z\to\ell\ell$ processes with one additional non-prompt lepton.  Three regions
are defined according to the number of jets in the event: one jet (\TLjone),
two jets (\TLjtwo), and at least three jets (\TLjthree).  For the four-lepton
channel, the third and fourth leading leptons are required to be nominal
leptons. The two leptons which are not part of the best $Z$ candidate
definition are required to have opposite charges. These ``other leptons'' are
used to define three regions, depending on whether they are different-flavour
(\FLDF), or same-flavour and their mass lies within a window of $10\,\GeV$
around the $Z$ boson mass (\FLSFZ) or their mass is outside this window
(\FLSFnoZ).

In each of the six regions the distribution of a dedicated
boosted-decision-tree discriminant, separating the \WVZ signal from the
dominant diboson background, is fed as input to the binned maximum-likelihood
fit to extract the signal.  For the three-lepton channels, 13, 15, and 12 input
variables are used for the \TLjone, \TLjtwo, and \TLjthree final states,
respectively, while for the four-lepton channels, six input variables are used
for each of the \FLDF, \FLSFZ and \FLSFnoZ final states.

Due to the required presence of nominal leptons in the three- and four-lepton
channels, backgrounds with a $Z$ boson and non-prompt leptons are reduced.  The
remaining backgrounds are dominated by processes with prompt leptons and thus
all backgrounds are estimated using simulation.  The $WZ+$jets and $Z$+jets
backgrounds are validated in a region defined in the same way as the \TLjone
region, with the exception that no requirement on \HT is applied, the
third-highest-\pt lepton is required to have a small transverse momentum
($10\,\GeV < \pt < 15\,\GeV$), and the invariant mass of the three leptons has
to be smaller than $150\,\GeV$.  Data and expectation agree in a dedicated
validation region.  The \ttZ background is determined in a region defined like
the \TLjthree region with the exception that no requirement on \HT is applied,
and at least four jets are required, of which at least two are $b$-tagged. This
region is included as a single-bin control region (CR) in the fit model.

\section{Signal extraction and combination} 

The \WWW, \WWZ and \WZZ regions are combined using the profile likelihood
method based on a simultaneous fit to distributions in the signal and
background control regions.  A total of eleven signal regions are considered:
four regions (\elel, \elmu, \muel, and \mum) for the \ssll channel, one region
($\mu e e$ and $e \mu \mu$ combined) for the \threelep channel, three regions
(\TLjone, \TLjtwo, and \TLjthree) for the \WVZ three-lepton channel, and three
regions (\FLDF, \FLSFZ, and \FLSFnoZ) for the \WVZ four-lepton channel.  One
control region is considered: the \ttZ control region described in
Section~\ref{s:wvz}.  The distributions used in the fit are the $m_{jj}$
distributions for the \ssll channel and the BDT distributions for the \WVZ
three-lepton and four-lepton channels. The number of selected events in the
\threelep channel and the \ttZ control region are each included as a single bin
in the fit. In total, 186 bins are used in the combined fit.  Correlations of
systematic uncertainties arising from common sources are maintained across
processes and channels.  

\begin{figure}[htbp]
\centering
\hfill
\subfloat[]{\includegraphics[width=.42\textwidth]{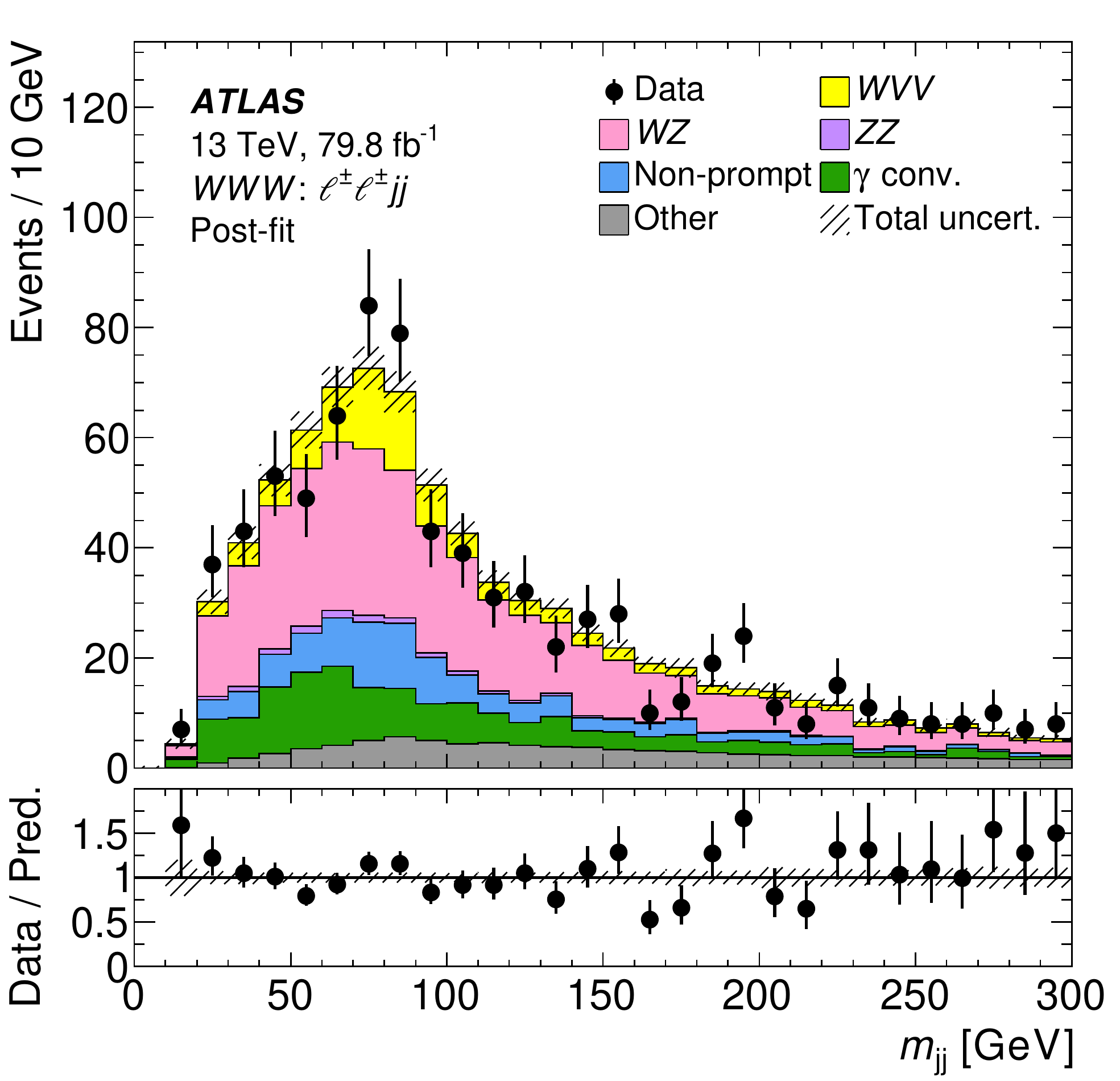}}
\hfill\hfill
\subfloat[]{\includegraphics[width=.42\textwidth]{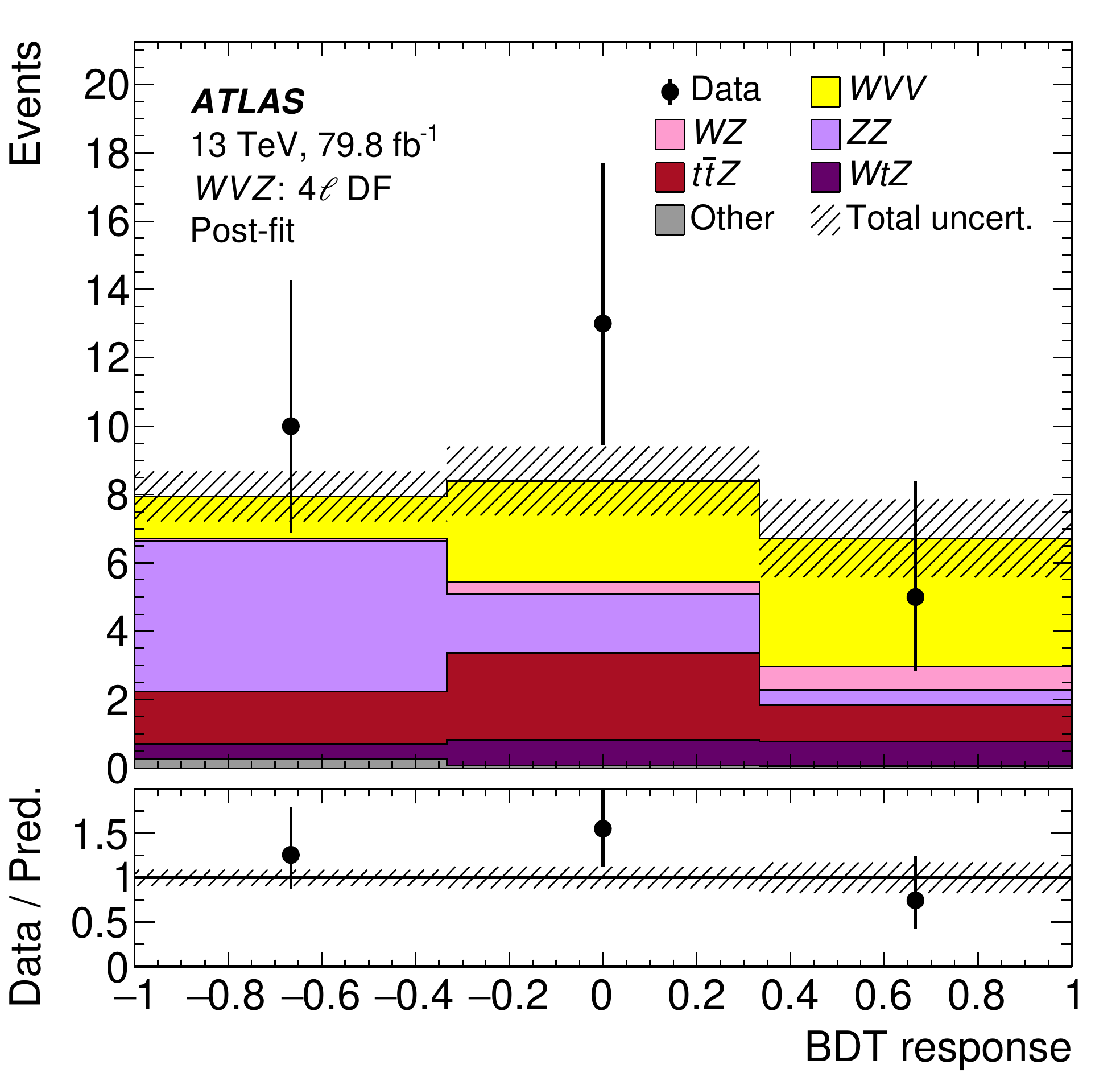}}
\hfill
\caption{\label{fig:www_wvz_post} Post-fit distribution of (a) $m_{jj}$ for the
\sswww analysis (\elel, \elmu, \muel, \mum combined) and (b) the BDT response
in the \FLDF channel for the $WVZ$ analysis~\cite{WVV}.  The uncertainty band
includes both statistical and systematic uncertainties as obtained by the fit.}
\end{figure}

The simultaneous fit model has the power to constrain the normalisations of the
dominant backgrounds from the $WZ$ and $ZZ$ processes at the $\sim$5\% level.
The contribution to the $WVV$ signal from $VH$ associated production is
$\sim$40\% in the $WWW$ fiducial regions and $\sim$30\% in the $WVZ$ fiducial
regions.  Figure~\ref{fig:www_wvz_post} shows the comparison between data and
post-fit prediction of the combined $m_{jj}$ distribution for the \lljj
channel, the number of selected events for the \lllv channel, and the BDT
output distributions in the \TLjtwo and \FLDF regions for the \WVZ analysis.
The \TLjtwo and \FLDF regions are chosen since they have the best sensitivity
among the three-lepton and four-lepton channels.  Data and predictions agree in
all distributions. 

The overall observed (expected) significance for $WVV$ production is found to
be 4.1$\sigma$ (3.1$\sigma$), constituting evidence for the production of three
massive vector bosons.  The combined best-fit signal strength for the $WVV$
process, obtained by the fit to the eleven signal regions and one control
region, is $\mu_{WVV} = \obsvvverr$ with respect to the SM prediction.  The
statistical uncertainty in the measured signal strength is $^{+0.25}_{-0.24}$
and the systematic uncertainty is $^{+0.30}_{-0.27}$.  The largest systematic
uncertainties come from uncertainties related to data-driven background
evaluations affecting the \WWW channels, from theoretical uncertainties related
to renormalisation and factorisation scale variations, and from experimental
uncertainties.

Fits are also performed separately in the $WWW$ and the $WVZ$ channels.  where
the other signal strength is fixed to its SM expectation.  For the fits of the
\WWW channels, an additional $WZ$ control region is used in the fit, helping to
constrain the overall normalisation of the $WZ$+jets background, which in the
combined fit is constrained by the \WVZ three-lepton signal regions.  The
observed (expected) significance is 3.2$\sigma$ (2.4$\sigma$) for $WWW$
production and 3.2$\sigma$ (2.0$\sigma$) for $WVZ$ production.  

\begin{table}[htbp]
\begin{center}
\begin{tabular}{l|cc}
\hline\hline 
\multirow{2}{*}{Decay channel} & \multicolumn{2}{c}{Significance}\\
 &     Observed       & Expected \\
\hline
$WWW$ combined          & 3.2$\sigma$ & 2.4$\sigma$ \\
\qquad \sswww                  & 4.0$\sigma$ & 1.7$\sigma$ \\
\qquad \lllwww                 & 1.0$\sigma$ & 2.0$\sigma$ \\ 
\hline
$WVZ$ combined          & 3.2$\sigma$ & 2.0$\sigma$ \\
\qquad \lllwvz                 & 0.5$\sigma$ & 1.0$\sigma$ \\
\qquad $\WVZ \to \ell \nu \ell \nu \ell \ell / qq \ell \ell \ell \ell$  & 3.5$\sigma$ & 1.8$\sigma$  \\ 
\hline
$WVV$ combined                    & 4.1$\sigma$ & 3.1$\sigma$ \\
\hline
\hline
\end{tabular}
\caption{Observed and expected significances with respect to the SM
background-only hypothesis for the four $WVV$ channels entering the fit. 
\label{tab:sensitivity}}
\end{center}
\end{table} 

\begin{figure}[htbp]
\centering
\hfill
\subfloat[]{\includegraphics[width=0.53\textwidth]{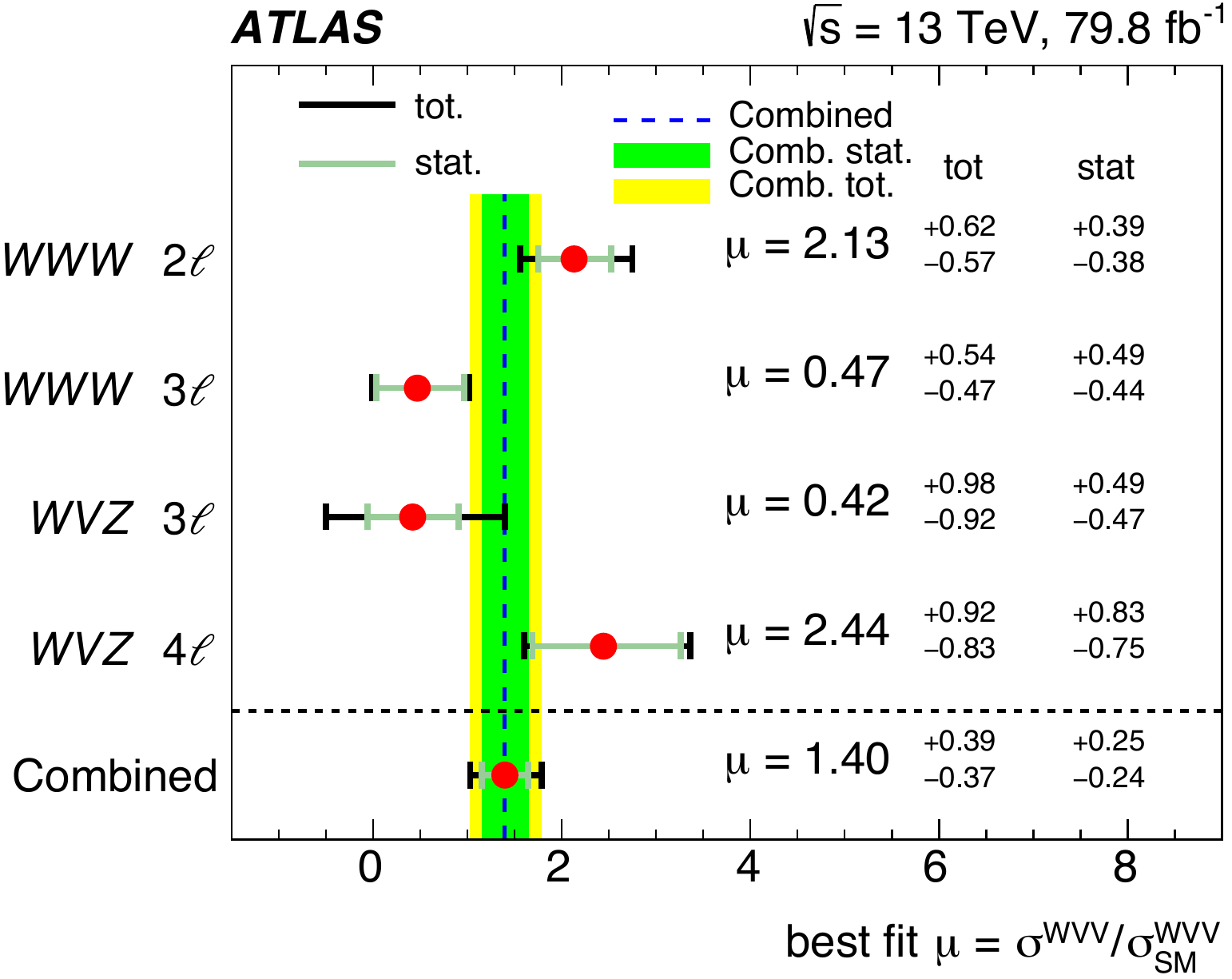}}
\hfill\hfill
\subfloat[]{\includegraphics[width=.44\textwidth]{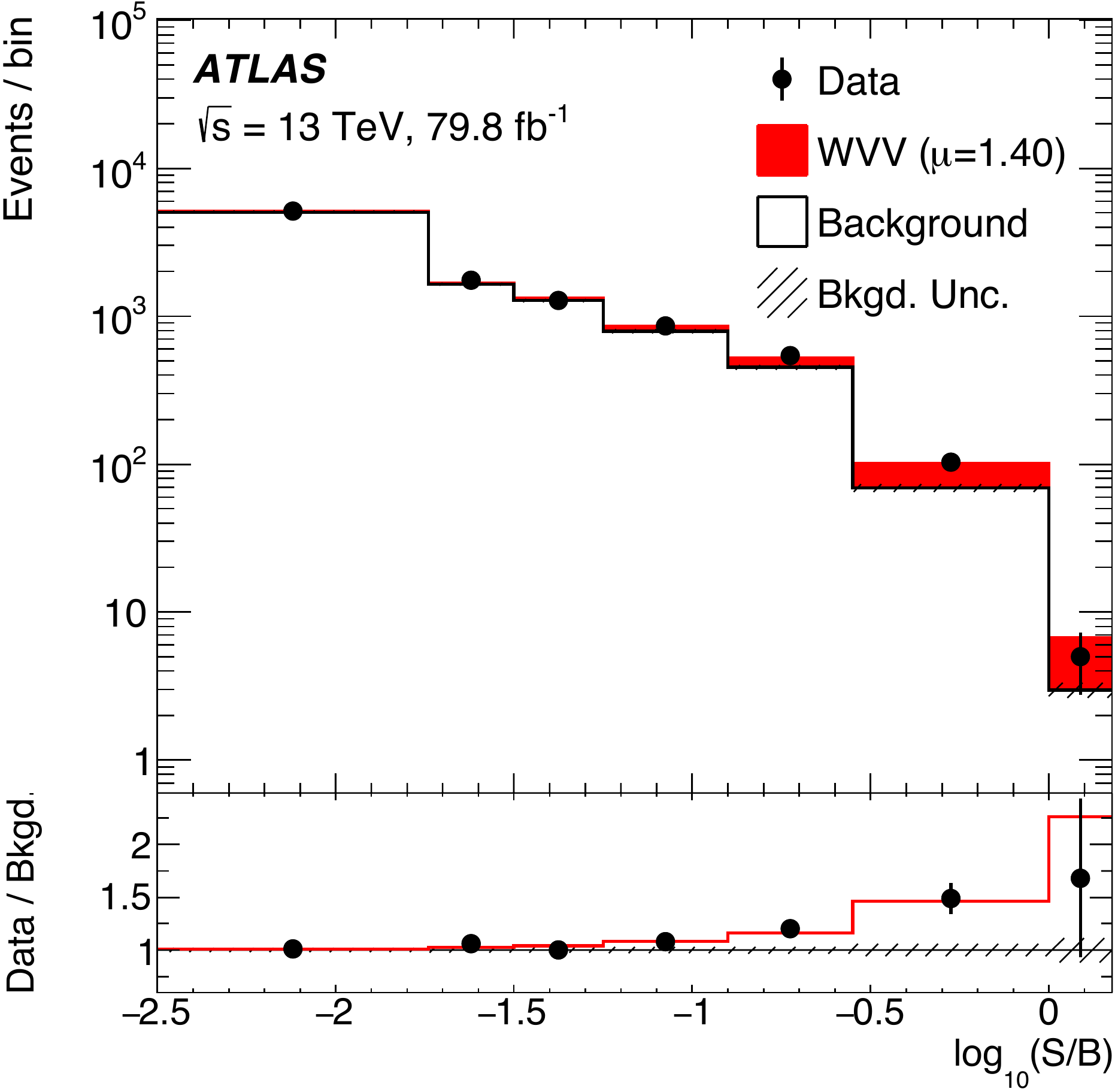}}
\hfill
\caption{(a) Extracted signal strengths $\mu$ for the four analysis regions and
for the combination.  (b) Event yields as a function of
$\log_{10}\left(S/B\right)$ for data, background $B$ and the signal $S$.
Events in all eleven signal regions are included. The background and signal
yields are shown after the global signal-plus-background fit.  The hatched band
corresponds to the systematic uncertainties, and the statistical uncertainties
are represented by the error bars on the data points.  The lower panel shows
the ratio of the data to the expected background estimated from the fit,
compared to the expected distribution including the signal (red line)~\cite{WVV}.
\label{fig:fig5}}
\end{figure}

Table~\ref{tab:sensitivity} and Figure~\ref{fig:fig5}(a) summarise the observed
and expected significances with respect to the background-only hypothesis and
the observed best-fit values of the signal strength for the individual and
combined fits. The measured signal strengths from the individual fits are
converted to inclusive cross-section measurements using the signal samples and
the central values of the theoretical predictions.  All uncertainties
determined in the fit are included in the conversion, except for the
normalisation uncertainty in the signal prediction.  The results are:
$\sigma_{WWW} =
0.65^{+0.16}_{-0.15}\,\textrm{(stat.)}\,^{+0.16}_{-0.14}\,\textrm{(syst.)}$ pb
and $\sigma_{WWZ} = 0.55 \pm
0.14\,\textrm{(stat.)}\,^{+0.15}_{-0.13}\,\textrm{(syst.)}$ pb.  For the
$\sigma_{WWZ}$ extraction, the $WZZ$ normalisation is fixed to the SM
expectation. The cross section of the latter is not reported, since there is
not enough sensitivity to this channel to quote a separate cross-section value.

Figure~\ref{fig:fig5}(b) shows the data, background and signal yields, where
the discriminant bins in all signal regions are combined into bins of
$\log_{10}(S/B)$, $S$ being the expected signal yield and $B$ the background
yield. The background and signal yields are shown after the global
signal-plus-background fit to the data.

\section{Summary} 

The search for the joint production of three massive vector bosons ($W$ or $Z$)
in proton--proton collisions using 79.8 fb$^{-1}$ of data at $\sqrt{s} =
13\,\TeV$ collected by the ATLAS detector at the LHC, was presented.  Events
with two, three or four reconstructed electrons and muons were analysed.  First
evidence for the production of three massive vector bosons has been observed
with a combined significance of 4.1 standard deviations, where the expectation
is 3.1 standard deviations.  The measured production cross sections are
$\sigma_{WWW} = 0.65^{+0.23}_{-0.21}$ pb, and $\sigma_{WWZ} =
0.55^{+0.21}_{-0.19}$ pb, in agreement with the Standard Model predictions.

\section*{Acknowledgements}

This work was partially funded by the European Research Council under the
European Union's Seventh Framework Programme ERC Consolidator Grant Agreement
n.~617185 (TopCoup) and by the German Federal Ministry of Education and
Research (BMBF) in FSP-103 under grant n.~05H15PDCAA.

\end{document}